\documentclass{article}
\usepackage{amssymb}

\usepackage{graphicx}
\usepackage{amsmath}

\input{tcilatex}

\begin{document}

{\huge Key Agreement and Authentication Schemes Using Non-Commutative
Semigroups}\medskip

M.M. Chowdhury\medskip

\underline{Abstract}\medskip

We give a new two-pass authentication scheme, which is a generalisation of
an authentication scheme of Sibert-Dehornoy-Girault based on the
Diffie-Hellman conjugacy problem. Compared to the above scheme, for some
parameters it is more efficient with respect to multiplications. We sketch a
proof that our authentication scheme is secure.\medskip

\underline{1. Introduction}\medskip

In recent years various cryptographic protocols using infinite non-abelian
groups have been proposed. For example the seminal algebraic key
establishment protocol given in \cite{AAG}, and Artin's braid groups have
been popular choices for such protocols. Braid groups are a popular choice
because they are not too complicated to work with and they are more
complicated than abelian groups. In particular the conjugacy problem in
braid groups is algorithmically difficult and hence gives a one-way function.

We give a new authentication scheme (by using equations (2) and (3) below
which form a main part of this paper), which is a generalisation of the
authentication scheme $I$ of SDG (Sibert-Dehornoy-Girault) given in \cite
{Sibert}. We do not claim that equations (2) and (3) are totally original;
for example a simpler version of equation (3) is used in the authentication
scheme in \cite{Stickel2004}. In this paper we refer to the authentication
scheme $I$ in \cite{Sibert} as the SDG scheme. Based on the proof of the SDG
scheme given in \cite{Sibert} we sketch a proof that our new authentication
scheme is a perfectly honest-verifier ZK interactive proof of knowledge of
the prover's secret. Two other provably secure schemes which are scheme $II$
and scheme $III$ are given \cite{Sibert} and they differ from our
authentication scheme because they are zero-knowledge in an theoretical
infinite framework, and they are iterated three pass schemes. Other
differences are that scheme $II$ is based on different hard problems
compared to our authentication scheme. We refer to \cite{Sibert} for further
details of these schemes. Two related authentication schemes were proposed
in \cite{Lal and Chaturvedi} and it was shown in \cite{Tsaban} one $``$%
authentication scheme$"$ in \cite{Lal and Chaturvedi} is totally insecure
and the other authentication scheme as shown in \cite{Tsaban} can be broken
by solving a specialisation of the decomposition problem defined in \cite
{Shpilrain et al} (see below). An authentication scheme is given in \cite
{Stickel2004} which has security based on a generalisation of the discrete
logarithm problem in non-abelian finite groups. The main difference of our
authentication scheme from the authentication scheme in \cite{Stickel2004}
and the unbroken scheme in \cite{Lal and Chaturvedi}, \cite{Tsaban} is that
it is based on our version of the Diffie-Hellman decomposition problem
defined below.\medskip

\underline{2. Hard Problems in Non-Abelian Groups}\medskip

We now define the following known hard problems. The notation $[I,J]=1$
(resp. $[I,J]\neq 1$) means that the subsets $I$ and $J$ of a semigroup $G$
commute (resp. do not commute). We may consider $G=B_{n}$ (the braid group
of index $n$) in the following known problems, because the problems are then
hard and by hard we mean there is no known algorithm to solve the problem
such that a cryptographic protocol based on the problem would be insecure
for practical use. When $G=B_{n}$ then WLOG the usual choices for $A$ and $B$
are the braid subgroups $LB_{n}$ and $UB_{n}$ (defined in section 3) as
these are the choices that are used in \cite{Ko et al} and \cite{Cheon et al}
but the choices of $A$ and $B$ may be different. For our protocol to be
secure at the very least $G$ should be non-commutative.\medskip \newline
The \textbf{DP} (Decomposition Problem) \cite{Shpilrain et al} is defined as
follows.\newline
\textit{Public Information}: $G$ is a semigroup, $A$ is a subset of $G$. $%
x,y\in G$ with $y=axb$.\newline
\textit{Secret information}: $a,b\in A$.\newline
\textit{Objective}: find elements $f,g\in A$ such that $fxg=y$.\medskip 
\newline
The definition of the $DP$ above generalises the definition of a less
general version of the $DP$ given in \cite{Cheon et al}, \cite{Ko} and \cite
{Shpilrain2 et al}. The less general version only differs from the above
definition of $DP$ because $G$ is a group and $A$ is a subgroup. In our
notation in all of this paper we omit the binary operation $\ast $ when
writing products so for example $f\ast x\ast g$ is understood to mean $fxg$.
We require that $\ast $ is efficiently computable.\medskip \newline
The \textbf{CSP} (Conjugacy Search Problem) \cite{AAG}, \cite{Ko} is defined
as follows.\newline
\textit{Public Information}: $G$ is a group. $x,y\in G$ with $y=f^{-1}xf$.%
\newline
\textit{Secret Information}: $f\in G$.\newline
\textit{Objective}: find an element $g\in G$ such that $g^{-1}xg=y$.\medskip 
\newline
The \textbf{DH-DP} (Diffie-Hellman Decomposition Problem) \cite{Cheon et al}%
, \cite{Ko} is defined as follows.\newline
\textit{Public Information}: $G$ is a group. $A,B$ are subgroups of $G$ with 
$[A,B]=1$. $x,y_{a},y_{b}\in G$ with $y_{a}=axb,$ $y_{b}=cxd$.\newline
\textit{Secret Information}: $a,b\in A$, $c,d\in B$.\newline
\textit{Objective}: find the element $cy_{a}d(=ay_{b}b=acxbd).$\medskip 
\newline
The \textbf{DH-CP} (Diffie-Hellman Conjugacy Problem) is the specialisation
of the DH-DP \cite{Cheon et al} with $a=b^{-1}$ and $c=d^{-1}$.\medskip 
\newline
We now re-define the DP and DH-DP above as used in our authentication
scheme. In the rest of this paper below the DP and DH-DP will mean their
re-definitions.\medskip \newline
The re-definition of the \textbf{DP} is as follows.\newline
\textit{Public Information}: $G$ is a semigroup. $A$, $B$ are subsets of $G$%
. $x,y\in G$ with $y=axb$.\newline
\textit{Secret Information}: $a\in A$, $b\in B$.\newline
\textit{Objective}: find elements $f\in A$, $g\in B$ such that $fxg=y$%
.\medskip \newline
The re-definition of the \textbf{DH-DP} is as follows.\newline
\textit{Public information}: $G$ is a semigroup. $A,B,C,D$ are subsets of $G$%
. $x,y_{a},y_{b}\in G$ with $y_{a}=axb$, $y_{b}=cxd$.\newline
\textit{Secret Information}: $a\in A$, $b\in B$, $c\in C$, $d\in D$.\newline
\textit{Objective}: find the element $cy_{a}d$ $(=ay_{b}b=acxbd)$. So if we
can find $f\in A$, $g\in B$ such that $fxg=y_{a}$ or $h\in C$, $i\in D$ such
that $hxi=y_{b}$ then this is sufficient to break our schemes which have
security based upon the DH-DP. \medskip

A variant of the above re-definition of the \textbf{DH-DP} which we refer to
as the \textbf{DH-DP' }on which the security of our variant protocols is
based upon is as follows.\newline
\textit{Public information}: $G$ is a semigroup. $A,B,C,D$ are subsets of $G$%
. $x,y_{a},y_{b}\in G$ with $y_{a}=axb$, $y_{b}=caxbd$.\newline
\textit{Secret Information}: $a\in A$, $b\in B$, $c\in C$, $d\in D$. ($a,b$
have an inverse).\newline
\textit{Objective}: find the element $a^{-1}y_{b}b^{-1}(=cxd,y_{a}=axb)$. So
if we can find $f\in A$, $g\in B$ $h\in C$, $i\in D$ such that $hxi=cxd$ or $%
f^{-1}y_{b}g^{-1}=cxd,$ $fxg=y_{a}$ then this is sufficient to break our
schemes which have security based upon the DH-DP'.\medskip

In all of this paper for our authentication scheme the DP, DH-DP and DH-DP'
are considered with commutativity conditions such as (2) and (3) defined
below. We assume the DP, DH-DP and DH-DP' are hard. The connection between
the DH-DP and the DP is similar to the one between the Diffie-Hellman
problem and the discrete logarithm problem. The DH-DP, DH-DP' is obviously
reducible to the DP, but we assume that it is as hard for general $G$. Hence
checking that the DP is hard for $x$ is supposed to ensure that the DH-DP
also is.

The security of the modified key exchange protocol on page 2 of \cite
{Shpilrain et al} is based on the DP with the additional condition that $%
[A,B]=1$ and its security is also based on the DH-DP.\medskip

\underline{3. The Sibert-Dehornoy-Girault Authentication Scheme}\medskip

All the details of implementation in braid groups for the SDG authentication
scheme $I$ are given in \cite{Sibert} so we do not reproduce them all here;
we restrict to the details we require. For $n\geq 2$, $B_{n}$ is defined to
be the group with the presentation with $n-1$ generators (plus the identity $%
e$), denoted $\sigma _{i}$ for $i=1,2,...,n-1$ and the defining
relationships 
\begin{equation}
\begin{array}[t]{ll}
\sigma _{i}\sigma _{j}=\sigma _{j}\sigma _{i} & |i-j|>1 \\ 
\sigma _{i}\sigma _{j}\sigma _{i}=\sigma _{j}\sigma _{i}\sigma _{j} & |i-j|=1
\end{array}
\tag{1}
\end{equation}
We refer the reader to any textbook about braids, for instance \cite
{Murasugi et al}; each element of $B_{n}$ has the geometrical interpretation
by an $n$-strand braid in the usual sense. This geometrical interpretation
is that any $n$-strand braid diagram can be first sliced into a
concatenation of elementary diagrams with one crossing each and then each
elementary diagram can be used to give an encoding of the braid diagram as a
word in one of the letters $\sigma _{i}$ or $\sigma _{i}^{-1}$. $\sigma _{i}$
is used for the diagram where the $i$th strand crosses under the $(i+1)$ st
one, and $\sigma _{i}^{-1}$ is used for the diagram where the $i$th strand
crosses over the $(i+1)$st one.

$LB_{n}$ and $UB_{n}$ are the two commuting subgroups of $B_{n}$ generated
by the Artin generators $\sigma _{1},...,\sigma _{\left\lfloor
n/2\right\rfloor -1}$ and $\sigma _{\left\lfloor n/2\right\rfloor
},...,\sigma _{n-1}$ \cite{Sibert}. The SDG authentication is as follows.
Following \cite{Sibert} we require that the DH-CP and CSP are hard in $G$ as
the security of the authentication is based on these problems, let $G=B_{n}$%
, $A=LB_{n}$, $B=UB_{n}$ in the DH-CP and CSP. Let $G^{^{\prime }}$ be a
non-abelian group, let $w\in G^{^{\prime }}$ be a publicly known word which
has a secret word as one of its factors (an attacker may attempt to recover
this secret word from $w$), let $S$ be a function which maps $w$ to an
equivalent word, $S$ is called a scrambling function, the security of the
protocol is based on the difficulty of recovering the above secret word from
the word $S(w)$ \cite{Sibert}. A choice for the scrambling function can be a
normal form for $w$ \cite{Sibert}. All braids are expressed in rewritten
form using the scrambling function such that the DH-CP and CSP are hard.

\begin{itemize}
\item  Phase 1. Key generation\newline
i) Choose a public $b\in B_{n}$.\newline
ii) $A$(lice) chooses a secret braid $s\in LB_{n}$, her private key; she
publishes $b^{^{\prime }}=sbs^{-1}$; the pair $(b,b^{^{\prime }})$ is the
public key.

\item  Phase 2. Authentication phase\newline
i) $B$(ob) chooses $r\in UB_{n}$ and sends the challenge $t=rbr^{-1}$ to $A$.%
\newline
ii) $A$ sends the response $y=h(sts^{-1})$ to $B$, and $B$ checks $%
y=h(rb^{^{\prime }}r^{-1})$.
\end{itemize}

$h$ is a fixed collision-free hash function from braids to sequences of 0's
and 1's or, possibly, to braids, for which this choice for $h$, $G$ must
have an efficient solution for the word problem for use in phase 2ii).

A proof that the above authentication scheme is a perfectly honest-verifier
ZK interactive proof of knowledge of $s$ is given in \cite{Sibert}. There is
a linear algebraic algorithm to solve the DH-CP but the attack is not
efficient enough to break the public-key cryptosystem with the proposed
parameters of \cite{Ko et al} in real time \cite{Cheon et al}. The SDG
scheme has a similar structure to the public-key cryptosystem in \cite{Ko et
al} because both algorithms are based on the DH-CP. Hence the parameters of
the SDG scheme are based on considerations of the parameters of the above
public-key cryptosystem \cite{Sibert}. Hence parameters can be chosen for
the SDG scheme so that it cannot be broken in real time by using the attack
on the DH-CP in \cite{Cheon et al}.\medskip

\underline{4. New Authentication Scheme}\medskip

The new authentication scheme is as follows. Let $G$ be a (infinite or
finite) non-commutative semigroup. We define the scrambling function as in
the SDG scheme, for our scheme, but with the modification that it is defined
over the semigroup $G$ instead of only a group. All elements in $G$ in this
section are rewritten using the scrambling function and parameters are
chosen such that the DH-DP and DP are hard, as the security of our
authentication scheme is based on these problems.

\begin{itemize}
\item  Phase 0. Initial setup\newline
i) $G$ is chosen and is publicly known. The users publicly agree on which
method first, second or third (described below) will be used to select the
subsets and publicly agree on which of the commutativity conditions (2) or
(3) will be used.\newline
A first method to select the parameters is to select publicly known subsets $%
L_{A}$, $L_{B}$, $R_{A}$, $R_{B}$ and $Z$ of $G$ are chosen for which either
property a) below is true or property b) below is true. Let $z\in Z$ with $z$
the publicly known element which is the value of $x$ in the definition of
the DH-DP used in the example of the DH-DP in our new authentication scheme.%
\newline
Following \cite{Shpilrain2 et al} let $g\in G$ for $G$ a group, $C_{G}(g)$
is the centraliser of $g$, we describe the modifications to the
authentication scheme (and these apply to the key agreement protocol
described below) to give two further methods to select the subgroups as
follows. Publicly known subsets or privately known $L_{A}$, $L_{B}$, $R_{A}$%
, $R_{B}$ and $Z$ of $G$ are chosen for which either property 2) below is
true or property 3) below for the second and third methods below.
\end{itemize}

Using the above first method the security of the protocol is based on the
DH-DP. We now give the two further methods to select the subsets which
result in two modifications to the protocol when the first method is used.
The second method to select the subgroups is $A$ chooses $(a_{1},a_{2})\in
G\times G$ and publishes the subgroups as a set of generators of the
centralisers $L_{B},R_{B},L_{B}\subseteq C_{G}(a_{1}),R_{B}\subseteq
C_{G}(a_{2}),L_{B}=\{\alpha _{1},...,\alpha _{k}\}$ etc. $B$ chooses $%
(b_{1},b_{2})\in L_{B}\times R_{B}$, and hence can compute $x$ below etc.
Following \cite{Shpilrain2 et al} there is no explicit indication of where
to select $a_{1},b_{1}$ and/or $a_{2},b_{2}$ from. Hence before attempting
something like a length based attack in this case the attacker has to
compute the centraliser of $L_{B}$ and/or $R_{B}$.\medskip

So a third method (this method is given for the key agreement protocol in 
\cite{Shpilrain2 et al} ) to select the subgroups is

$A$ chooses $L_{A}=G,a_{1}\in G$, and publishes $L_{B}\subseteq
C_{G}(a_{1}),L_{B}=\{\alpha _{1},...,\alpha _{k}\}$,

$B$ chooses $L_{B}=G,b_{2}\in G$, and publishes $R_{A}\subseteq
C_{G}(b_{2}),R_{A}=\{\beta _{1},...,\beta _{k}\}$,

Hence $A$ chooses $(a_{1},a_{2})\in G\times C_{G}(b_{2})$ and publishes the
subgroup(s) as a set of generators of the centralisers $B$ chooses $%
(b_{1},b_{2})\in C_{G}(a_{1})\times G$, and hence can compute $x$ etc. Again
there is no explicit indication of where to select $a_{1}$ and/or $b_{2}$
from. Hence before attempting a length based attack in this case the
attacker has to compute the centraliser of $L_{B}$ and/or $R_{B}$.

The above three methods are examples of the following method, in general if
a user $A$ or $B$ selects an element $a_{i}$ or $b_{i}$ respectively as
their secret element (which multiplies the public element $z$ at the left
WLOG) key then the other user selects an element of the $C_{G}(a_{i})$ or $%
C_{G}(b_{i})$ respectively as their secret key for the corresponding secret
commuting element to left multiply WLOG, and in general an attacker has no
explicit indication of where to select $a_{i}$ and/or $b_{i}$ from. A
potential disadvantage of using the above third method to select the
subgroups is one user chooses the other users subgroups and this may aid the
user who has selected the subgroup to find the others users secret key, and
this secret key may be of use in say another attack. Hence before attempting
a length based attack in this case the attacker has to compute the
centraliser of $C_{G}(a_{i})$ and/or $C_{G}(b_{i})$. Note if the subgroups
are selected in this in the second or third way (using centraliser
computations) then its security is based on a variant of the DP, DH-DP and
the difficulty of computing centralisers, for example if the third method
above is used to select the subsets then following the attack given in \cite
{Shpilrain2 et al} the security of the protocol to find $A$ or $B^{\prime }s$
private key may be found as follows\medskip

Attack on $A$'s Key. Find an element $a_{1}^{\prime }$ which commutes with
every element of the subgroup $L_{B}$ and an element $a_{2}^{\prime }\in
R_{B}$ such that $z^{\prime }=a_{1}^{\prime }za_{2}^{\prime }$ where $%
a_{1}^{\prime }za_{2}^{\prime }$ above may be rewritten using a normal form.
The pair $(a_{1}^{\prime },a_{2}^{\prime })$ is equivalent to the pair $%
(a_{1},a_{2}),$ because $a_{1}^{\prime }za_{2}^{\prime }=a_{1}za_{2}$ this
means an attacker can authenticate as Alice. The attack applies to the key
exchange protocol below with the modification $K_{A}=a_{1}^{\prime
}za_{2}^{\prime }$ instead of $z^{\prime }=a_{1}^{\prime }za_{2}^{\prime }$
this gives a equivalent secret key for $A$ used to get the common secret key.

Attack on $B$'s Key in the key exchange protocol in the section below. Find
an element $b_{2}^{\prime }$ which commutes with every element of the
subgroup $R_{A}$ and an element $b_{1}^{\prime }\in L_{B}$ such that $%
K_{B}=b_{1}^{\prime }zb_{2}^{\prime }$ where $a_{1}^{\prime }za_{2}^{\prime
} $ above may be rewritten using a normal form. The pair $(b_{1}^{\prime
},b_{2}^{\prime })$ is equivalent to the pair $(b_{1},b_{2}),$ because $%
b_{1}^{\prime }zb_{2}^{\prime }=b_{1}zb_{2}$ this means an attacker can find
the common secret key when this set up is used as part of a key exchange
algorithm as described below.

Then (following \cite{Shpilrain2 et al}) the most obvious way to recover Bob
private key (the attack for $A$ is similar)\medskip

B1. Compute the centraliser of $R_{A},R_{A}\subseteq C_{G}(b_{2})$.

B2. Solve the search version of the membership problem in the double coset $%
<L_{B}>\cdot z\cdot C_{G}(R_{A})$.\medskip

So for the protocol to be secure we want both the above problems to be
computationally hard, for the problem B2 to be hard it is required the
centraliser $C_{G}(L_{A})$ should be large enough to resist a brute force
attack. The key exchange protocol in section 5 has security based upon the
above problem. The above attack can be used to attack the authentication
scheme or recover $B$'s secret elements with the modification $%
x=b_{1}^{\prime }z^{\prime }b_{2}^{\prime }$ instead of $K_{B}=b_{1}^{\prime
}zb_{2}^{\prime }$ ($x$ is used instead of $K_{B}$ and $z^{\prime }$ is used
instead of $z$ etc.).

The attacks above are considered with commutativity condition for the DH-DP
such as (2) and (3) below. A similar attack is discussed when the second
method is used to choose the subsets. Hence the platform group $G$ should
satisfy the requirements given in the security analysis in section 6.

a) If $z\neq e$ we require the following conditions 
\begin{equation}
\begin{array}[t]{ll}
\lbrack L_{A},L_{B}]=1, & [R_{A},R_{B}]=1, \\ 
\lbrack L_{B},Z]\neq 1, & [L_{A},Z]\neq 1, \\ 
\lbrack R_{B},Z]\neq 1, & [R_{A},Z]\neq 1, \\ 
\lbrack L_{A},R_{A}]\neq 1, & [L_{B},R_{B}]\neq 1.
\end{array}
\tag{2}
\end{equation}
All the above conditions for $z\neq e$ can arise by generalising from
properties of subgroups used in either the SDG scheme or CKLHC scheme for
example the second and third conditions in (2) arise from the observations
that in general $[LB_{n},B_{n}]\neq 1,[LB_{n},UB_{n}]=1$.\medskip \newline
b) If $z=e$ we require the following conditions 
\begin{equation}
\begin{array}[t]{ll}
\lbrack L_{A},L_{B}]=1, & [R_{A},R_{B}]=1, \\ 
\lbrack L_{A},R_{A}]\neq 1, & [L_{B},R_{B}]\neq 1, \\ 
\lbrack L_{B},R_{A}]\neq 1, & [L_{A},R_{B}]\neq 1.
\end{array}
\tag{3}
\end{equation}
We were unable to show that the DH-DP using conditions (2) or (3) was easy.
Hence we assume that the DH-DP is hard with condition with the above
conditions. Note condition (3) is condition (2) but with conditions of the
form of a subset not commuting with $Z$ omitted (because $z=e$) and the
additional conditions $[L_{B},R_{A}]\neq 1,[L_{A},R_{B}]\neq 1$. The above
additional conditions are required so the DH-DP is not easy and hence our
authentication scheme is secure.

\begin{itemize}
\item  Phase 1. Key generation\newline
i) Choose a public $z\in Z$.\newline
ii) $A$ chooses secret elements $a_{1}\in L_{A}$, $a_{2}\in R_{A}$, her
private key; she publishes $z^{^{\prime }}=a_{1}za_{2}$; the pair $%
(z,z^{^{\prime }})$ is the public key.

\item  Phase 2. Authentication phase\newline
i) $B$ chooses $b_{1}\in L_{B}$, $b_{2}\in R_{B}$ and sends the challenge $%
x=b_{1}zb_{2}$ to $A$.\newline
ii) $A$ sends the response $w=H(a_{1}xa_{2})$ to $B$, and $B$ checks $%
w=H(b_{1}z^{^{\prime }}b_{2})$.
\end{itemize}

$H$ is a fixed collision-free hash function from elements of $G$ to
sequences of 0$^{^{\prime }}$s and 1$^{^{\prime }}$s or, possibly, to
elements of $G$, for which this choice of $H$, $G$ must have an efficient
solution for the word problem for use in phase 2ii)\medskip

\underline{Proposition 4.1}\medskip

Our Authentication Scheme is a perfectly honest-verifier ZK interactive
proof of knowledge of $a_{1}$ and $a_{2}$.\medskip \newline
\textbf{Proof.}\textit{\ (Sketch) Completeness}. Assume that, at step 2(ii) $%
w^{^{\prime }}$ is sent by $A$. Then $B$ accepts $A$'s key if and only if $%
w^{^{\prime }}=H(b_{1}z^{^{\prime }}b_{2})$ which is equivalent to 
\begin{equation}
w^{^{\prime }}=H(b_{1}(a_{1}za_{2})b_{2}).  \tag{4}
\end{equation}

By hypothesis $a_{1}\in L_{A}$, $a_{2}\in R_{A}$, $b_{1}\in L_{B}$, $%
b_{2}\in R_{B}$, so $b_{1}a_{1}=a_{1}b_{1},b_{2}a_{2}=a_{2}b_{2}$ holds and
(4) is equivalent to $w=H(a_{1}(b_{1}zb_{2})a_{2})$ i.e. $w^{^{\prime }}=w$.

\textit{Soundness}. Assume cheater $A^{^{\prime }}$ is accepted with
non-negligible probability. This means $A^{^{\prime }}$ can compute $%
H(b_{1}z^{^{\prime }}b_{2})$ with non-negligible probability. Since $H$ is
supposed to be an ideal hash function, this means that $A^{^{\prime }}$ can
compute the element $q$ satisfying $H(q)=H(b_{1}z^{^{\prime }}b_{2})$ with
non-negligible probability and this is because of two possibilities. The
first possibility is that $q=b_{1}z^{^{\prime }}b_{2}$ which contradicts the
hypothesis that the DH-DP is hard. The second possibility is $q\neq
b_{1}z^{^{\prime }}b_{2}$ which means that $A^{^{\prime }}$ and $B$ are able
to find a collision for $H$ which contradicts that $H$ is a collision free
hash function.

\textit{Honest-verifier zero knowledge}. Consider the probabilistic Turing
machine defined as follows: it chooses random elements $b_{1}$ and $b_{2}$
using the same drawing as the honest verifier, and outputs the instances $%
(b_{1},b_{2},H(b_{1}z^{^{\prime }}b_{2}))$. So the instances generated by
this simulator follow the same probability distribution as the ones
generated by the interactive pair $(A,B)$. $\Box $\medskip

\underline{4.1 Comparison of Our Authentication Scheme with the}

\underline{Sibert-Dehornoy-Girault Authentication Scheme}\medskip

The generalisation (not the variants) specialises to the SDG scheme (hence
our authentication scheme can be as secure as the SDG authentication scheme)
with the parameters $H=h$, $G=B_{n},Z=B_{n}$, $L_{A}=R_{A}=LB_{n},$ $%
L_{B}=R_{B}=UB_{n}$, $b_{1}=r$, $b_{2}=r^{-1}$, $a_{1}=s$, $a_{2}=s^{-1}$, $%
z=b$, $z^{^{\prime }}=b^{^{\prime }}$, $x=t$ , $w=y$, the first method is
used to select the subsets and conditions (2) (or property a) is true) are
used.

If $z=e$ then this implies the following. Condition (3) do not allow the
subgroups used in the SDG scheme to be used in our authentication scheme,
because if these subgroups are used as the incorrect choices for $L_{A}$, $%
L_{B}$, $R_{A}$ and $R_{B}$ in our authentication scheme then it is easy to
see the DH-DP is easy and hence our authentication scheme is insecure. There
is more control of the public parameters (the choices of the subsets)
compared to the SDG scheme and this may be useful for selecting secure
public keys. Compared to the SDG scheme, potentially our authentication
scheme requires less memory and fewer multiplications to compute the
challenge of $B$ and the public key of $A$. This is because the identity
element may take little memory (compared to $b\neq e$) to represent
depending on the representation of $G$. Alternatively we can omit the
implementation of using $Z$ completely because $x$ and $z^{^{\prime }}$ can
be computed using two multiplications.\medskip

\underline{4.2 A Variant of the Authentication Scheme}\medskip

A variant of the above authentication scheme is as follows\medskip

\begin{itemize}
\item  Phase 0. Initial Setup. The phase 0 is the same as phase 0 of the
authentication scheme in section 4.

\item  Phase 1. Key generation\newline
i) Choose a public $z\in Z$.\newline
ii) $A$ chooses (invertible) secret elements $a_{1}\in L_{A}$, $a_{2}\in
R_{A}$, her private key; she publishes $z^{^{\prime }}=a_{1}za_{2}$; the
pair $(z,z^{^{\prime }})$ is the public key.

\item  Phase 2. Authentication phase\newline
i) $B$ chooses $b_{1}\in L_{B}$, $b_{2}\in R_{B}$ and sends the challenge $%
x=b_{1}z^{\prime }b_{2}$ to $A$.\newline
ii) $A$ sends the response $w=H(a_{1}^{-1}xa_{2}^{-1})$ to $B$, and $B$
checks $w=H(b_{1}zb_{2})$, if the check is true he accepts if the check is
false he rejects.\medskip
\end{itemize}

$H$ is a fixed collision-free hash function from elements of $G$ to
sequences of 0$^{^{\prime }}$s and 1$^{^{\prime }}$s or, possibly, to
elements of $G$, for which this choice of $H$, $G$ must have an efficient
solution for the word problem for use in phase 2ii). The above variant
authentication scheme has security based on the DH-DP'.

The above variant authentication scheme specialises to the authentication
scheme in \cite{Stickel2004}. The above variant protocol specialises to the
authentication scheme in \cite{Stickel2004} with the parameters, conditions
(3) are used, $G$ a finite non-abelian group, Bob is user $B$ in our
protocol, Alice is user $A$ in our protocol $z=identity$ $element$, $%
L_{A},R_{B}$ are publicly known and is generated by $a$, $R_{B}$ is
generated by $b$, $L_{B}=L_{A},R_{B}=R_{A}$ $B$ selects the secret element
which is depends on the secret exponents $0<r<n_{1},0<s<n_{2}$, $a^{r}\in
L_{A},b^{s}\in R_{B}$, $A$ selects the secret element which is depends on
the secret exponents $0<v<n_{1},0<w<n_{2}$, $a^{v}\in L_{A},b^{w}\in R_{B}$,
so the common secret key is $z^{^{\prime }}=a^{v}b^{w}$, where the notation $%
e,f$,is used in \cite{Stickel2004}.\medskip

\underline{5. New Key Agreement Protocol}\medskip

The setup for the authentication protocols in the above section can be used
for key agreement as follows.\medskip

\begin{itemize}
\item  Phase 0. Initial Setup. The phase 0 is the same as phase 0 of the
authentication scheme in section 4.

\item  Choose $z\in G$. Phase 1.\newline
ii) $A$(lice) chooses a secret elements $a_{1}\in L_{A}$, $a_{2}\in R_{A}$,
her private key; she publishes $K_{A}=a_{1}za_{2}$; the pair $(z,K_{A})$ is
the public key.\newline
i) $B$(ob) chooses a secret elements $b_{1}\in L_{B}$, $b_{2}\in R_{B}$, her
private key; she publishes $K_{B}=b_{1}zb_{2}$; the pair $(z,K_{B})$ is the
public key.\newline
iii) $A$ and $B$ can compute the common shared secret key $\kappa $ as $%
\kappa =a_{1}K_{B}a_{2}$ and $\kappa =b_{1}K_{A}b_{2}$ respectively.
Optionally the alternative computation $\kappa =h(a_{1}K_{B}a_{2})$ and $%
\kappa =h(b_{1}K_{A}b_{2})$ can be done.\newline
$\medskip $
\end{itemize}

$h$ is a fixed collision-free hash function from braids to sequences of 0's
and 1's or, possibly, to braids, for which this choice for $h$.\ Again the
above protocol is considered with the commutativity conditions 2 or 3.\ Note
the elements $K_{A}$ and $K_{B}$ are rewritten for example a normal form to
make the protocol secure.

Because phase 0 is the same as the authentication scheme again its security
is based on a variant of the DP, DH-DP and the difficulty of computing
centralisers.\medskip

Our protocol specialises to the CKLHC protocol in \cite{Cha et al} is the
above protocol with the parameters $G$\ the braid group, $A$ and $B$
commuting defined by $LB_{n}$ and $UB_{n}$ are the two commuting subgroups
of $B_{n}$ generated by the Artin generators $\sigma _{1},...,\sigma
_{\left\lfloor n/2\right\rfloor -1}$ and $\sigma _{\left\lfloor
n/2\right\rfloor },...,\sigma _{n-1},$ the first method in phase 0 is used
and the DH-DP uses the condition (2). The publicly transmitted information
is rewritten using the left canonical form. The KLCHKP protocol \cite{Ko et
al} is the specialisation of the CKLHC protocol with $a_{2}=a_{1}^{-1}$ and $%
b_{2}=b_{1}^{-1}$ and hence our protocol generalises the KLCHKP protocol.

Our protocol specialises to the key agreement protocol given on page two of 
\cite{Shpilrain et al} with the parameters $L_{A}=R_{B}$, $L_{B}=R_{A}$, $%
[L_{A},L_{B}]=1,$ and commutativity condition (2), $G$ is a semi-group or $G$
is the Thompson group, the first method in phase 0 is used and hence our
protocol generalise the protocol given in \cite{Shpilrain et al}. Using the
notation used in \cite{Shpilrain et al} this is $L_{A}=A,L_{B}=B,z=w$. So
the left and right secrets are taken from different subgroups the protocol.

Our protocol specialises to the key agreement protocol given in \cite
{Shpilrain2 et al} when $G$ is a group, $A$ and $B$ are subgroups in the
parameters of the DH-DP, using the third method in phase 0 to select the
subsets/ modify the key agreement protocol described above, commutativity
condition 2, $L_{B}\subseteq C_{G}(a_{1}),L_{B}=\{\alpha _{1},...,\alpha
_{k}\},R_{A}\subseteq C_{G}(b_{2}),R_{A}=\{\beta _{1},...,\beta _{k}\},K_{A}$
and $K_{B}$ are rewritten using a normal form, using the notation given in 
\cite{Shpilrain2 et al} this is $K_{A}=P_{A},K_{B}=P_{B},B=L_{B},A=R_{A}.$

Hence we get new variants (which use less multiplications) of the protocols 
\cite{Shpilrain et al}, \cite{Shpilrain2 et al}, \cite{Ko et al},\cite{Cha
et al} if we consider the parameters that our key agreement protocol above
specialises to the protocols above but using condition (3) instead of
conditions (2), for example if we consider the second method in phase 0 to
modify the protocol with the parameters above to specialise, then compared
to using the third method in phase 0 the advantage of this protocol is that
user $A$ does not have to wait for $B$ to select the subgroup and so for
example can create the certificate to prove that the public key of $A$
belongs to $A$.\medskip

\underline{5.1 Variant of Key Exchange}\medskip

A variant of the key exchange is as follows\medskip

The above setup can be used for key agreement as follows the details are as
follows.

\begin{itemize}
\item  Phase 0. Initial Setup. The phase 0 is the same as phase 0 of the
authentication scheme in section 4.

\item  Choose $z\in G$.\newline
ii) $A$(lice) chooses a secret invertible elements $a_{1}\in L_{A}$, $%
a_{2}\in R_{A}$, her private key; she publishes $K_{A}=a_{1}za_{2}$; the
pair $(z,K_{A})$ is the public key.\newline
i) $B$(ob) chooses a secret braid $b_{1}\in L_{B}$, $b_{2}\in R_{B}$, his
private key; he publishes $K_{B}=b_{1}K_{A}b_{2}$; the pair $(K_{A},K_{B})$
is the public key.\newline
iii) $A$ and $B$ can compute the common shared secret key $\kappa $ as $%
\kappa =a_{1}^{-1}K_{B}a_{2}^{-1}$ and $\kappa =b_{1}zb_{2}$ respectively.
Optionally the alternative computation $\kappa =h(a_{1}^{-1}K_{B}a_{2}^{-1})$
and $\kappa =h(b_{1}K_{A}b_{2})$ can be done.\medskip
\end{itemize}

$h$ is a fixed collision-free hash function from braids to sequences of 0's
and 1's or, possibly, to braids, for which this choice for $h$. Again the
above protocol is considered with the commutativity conditions 2 or 3. Note
the elements $K_{A}$ and $K_{B}$ are rewritten for example a normal form to
make the protocol secure.

Again because phase 0 is the same as the authentication scheme we have
sketched its security is based on a variant of the DP, DH-DP' and/or the
difficulty of computing centralisers.

The above variant protocol specialises to the key exchange in \cite{Stickel}
with the parameters, conditions (2) or (3) are used, $G$ a finite
non-abelian group, Bob is user $B$ in our protocol, Alice is user $A$ in our
protocol, $z=e$, $L_{A}$ is publicly known and is generated by $a$, $R_{B}$
is generated by $b$, $L_{B}=L_{A},R_{B}=R_{A}$ $B$ selects the secret
element which is depends on the secret exponents $0<r<n_{1},0<s<n_{2}$, $%
a^{r}\in L_{A},b^{s}\in R_{B}$, $A$ selects the secret element which is
depends on the secret exponents $0<v<n_{1},0<w<n_{2}$, $a^{v}\in
L_{A},b^{w}\in R_{B}$, so the common secret key is $\kappa =f=a^{v}b^{w}$,
where the notation $e,f$ is used in \cite{Stickel}.\medskip

\underline{5.1.1 Groups With No Efficient Normal Form}\medskip

If there is no efficient algorithm for a canonical form in $G$ then the
secret elements such as Alice's secret key must be disguised in another way
such as using a scrambling function. The algorithm in \cite{AAG} for when $G$
is a group, and there is no efficient normal form in $G$ but there is an
efficient algorithm for the word problem in $G$, can be used to find a
common shared secret key can be found as follows using the algorithm in \cite
{AAG} which is. Fix $b$ as $b=0$ or $b=1$.

1. $B$ sends a rewritten form of $\kappa $ which is $r$ for $r=\kappa $ or a
random word for $r,r\neq \kappa $.

2. User $A$ checks if $k=r$ then this determines the bit $b$, otherwise the
bit is $1-b$.

3. The steps 1 and 2 are repeated $m$ times so an $m$ bit key is exchanged.

As stated in \cite{AAG} the protocol is probabilistic and slower compared to
using a canonical form.\medskip

\underline{6. Security Analysis}\medskip

The attacks below are considered with commutativity condition for the DH-DP
such as (2) and (3) below.\medskip

\underline{6.1 Attacks When Second Method is used to Choose Subsets }

\underline{in Authentication Scheme and Key Exchange Scheme}\medskip

If the second method above is used to select the subsets then following the
attack given in \cite{Shpilrain2 et al} the security of the protocol to find 
$A$ or $B^{\prime }s$ private key may be found as follows\medskip

Attack on $A$'s Key. Find an element $a_{1}^{\prime }$ which commutes with
every element of the subgroup $L_{B}$ and an element $a_{2}^{\prime }\in
R_{B}$ which commutes with every element of the subgroup $R_{B}$ such that $%
z^{\prime }=a_{1}^{\prime }za_{2}^{\prime }$ where $a_{1}^{\prime
}za_{2}^{\prime }$ above may be rewritten using a normal form. The pair $%
(a_{1}^{\prime },a_{2}^{\prime })$ is equivalent to the pair $(a_{1},a_{2}),$
because $a_{1}^{\prime }za_{2}^{\prime }=a_{1}za_{2}$ this means an attacker
can authenticate as Alice. The attack applies to the key exchange protocol
(when the second method is used to choose the subsets) with the modification 
$K_{A}=a_{1}^{\prime }za_{2}^{\prime }$ instead of $z^{\prime
}=a_{1}^{\prime }za_{2}^{\prime }$ this gives a equivalent secret key for $A$
used to get the common secret key.

Then (following \cite{Shpilrain2 et al}) the most obvious way to do the
above attack Alice's private key\medskip

A1. Compute the centraliser of $R_{B},R_{B}\subseteq C_{G}(a_{2})$ and
compute the centraliser of $L_{B},L_{B}\subseteq C_{G}(a_{1})$

A2. Solve the search version of the membership problem in the double coset $%
<C_{G}(L_{B})>\cdot z\cdot <C_{G}(R_{B})>$.\medskip

So for the protocol to be secure we want both the above problems to be
computationally hard, for the problem A1 to be hard it is required both the
centralisers $C_{G}(L_{B})$ and $C_{G}(R_{B})\ $be large enough to resist a
brute force type attack. The above attack can be used to attack the
authentication scheme or recover $A$'s secret elements with the modification 
$z^{^{\prime }}=a_{1}^{\prime }za_{2}^{\prime }$ instead of $%
K_{A}=a_{1}^{\prime }za_{2}^{\prime }$ ($z^{\prime }$ is used instead of $%
K_{A}$ etc.) so this can be used to impersonate $A$.

Attack on $B$'s Key in the key exchange protocol. Find an element $%
b_{1}^{\prime }$ $\in L_{B}$ and an element $b_{2}^{\prime }\in R_{B}$ such
that $K_{B}=b_{1}^{\prime }zb_{2}^{\prime }$ where $b_{1}^{\prime
}zb_{2}^{\prime }$ above may be rewritten using a normal form. The pair $%
(b_{1}^{\prime },b_{2}^{\prime })$ is equivalent to the pair $(b_{1},b_{2}),$
because $b_{1}^{\prime }zb_{2}^{\prime }=b_{1}zb_{2}$ this means an attacker
can find the common secret key when this set up is used as part of a key
exchange algorithm as described below.

Then (following \cite{Shpilrain2 et al}) the most obvious way to do the
above attack, Bob private key \medskip

B1. Solve the search version of the membership problem in the double coset $%
<L_{B}>\cdot z\cdot <R_{B}>$.\medskip

So for the protocol to be secure we want the above problem to be
computationally hard, for the problem B1 to be hard it is required that the
elements cannot of $L_{B}$, $R_{B}$ all be tested (for$\ $an equivalent key
pair $b_{1}^{\prime },b_{2}^{\prime }$) so to resist a brute force attack
the possible values for $b_{1}^{\prime },b_{2}^{\prime }$ should be large
enough. The above attack can be used to attack the authentication scheme or
recover $B$'s secret elements with the modification $x^{\prime
}=b_{1}^{\prime }z^{\prime }b_{2}^{\prime }$ instead of $K_{B}=b_{1}^{\prime
}zb_{2}^{\prime }$ ($x^{\prime }$ is used instead of $K_{B}$ and $z^{\prime
} $ is used instead of $z$ etc.) so this can be used to impersonate $A$%
.\medskip

\underline{6.2 Attack When Second Method is used to Choose Subsets in the}

\underline{ Variant Authentication Scheme and Variant Key Exchange Scheme}%
\medskip

If the second method above is used to select the subsets then following
variant of the attack given in \cite{Shpilrain2 et al} the security of the
protocol to find $A$ or $B^{\prime }s$ private key may be found as
follows\medskip

Attack on $A$'s Key. Find an element $a_{1}^{\prime }$ which commutes with
every element of the subgroup $L_{B}$ and an element $a_{2}^{\prime }\in
R_{B}$ which commutes with every element of the subgroup $R_{B}$ such that $%
z^{\prime }=a_{1}^{\prime }za_{2}^{\prime }$ where $a_{1}^{\prime
}za_{2}^{\prime }$ above may be rewritten using a normal form and $%
a_{1}^{\prime },a_{2}^{\prime }$ are both invertible elements.

The pair $(a_{1}^{\prime },a_{2}^{\prime })$ is equivalent to the pair $%
(a_{1},a_{2}),$ because $a_{1}^{\prime }za_{2}^{\prime }=a_{1}za_{2}$ this
means an attacker can authenticate as Alice because the attacker can compute 
$a_{1}^{-1}xa_{2}^{-1}$. The above attack applies to the key exchange
protocol with the modification $K_{A}=a_{1}^{\prime }za_{2}^{\prime }$
instead of $z^{\prime }=a_{1}^{\prime }za_{2}^{\prime }$ this gives an
equivalent secret key for $A$ used to get the common secret key which can be
computed as $\kappa =a_{1}^{-1}K_{B}a_{2}^{-1}$.\smallskip\ (Another attack
call this attack B, is to find elements $a_{1}^{\prime },a_{2}^{\prime }$
such that $a_{1}^{\prime -1}z^{\prime }a_{2}^{\prime -1}=z$ , $a_{1}^{\prime
},a_{2}^{\prime }$ can be used instead of $a_{1},a_{2}$ to get the common
secret key, the attack is similar for the key agreement protocol with $K_{A}$
used in place of $z^{^{\prime }}$).

Then (following \cite{Shpilrain2 et al}) the most obvious way to do the
above attack Alice's private key\medskip

A1. Compute the centraliser of $R_{B},R_{B}\subseteq C_{G}(a_{2})$ and
compute the centraliser of $L_{B},L_{B}\subseteq C_{G}(a_{1})$

A2. Solve the search version of the membership problem in the double coset $%
<C_{G}(L_{B})>\cdot z\cdot <C_{G}(R_{B})>$ and/or $<C_{G}(L_{B})>\cdot
K_{B}\cdot <C_{G}(R_{B})>$ .\medskip

(For attack B we do the search $<C_{G}(L_{B})>\cdot z^{^{\prime }}\cdot
<C_{G}(R_{B})>$ $\ $there is a variant of attack B when the third method is
used to choose the subsets doing a search of the form $<L_{B}>\cdot
z^{^{\prime }}\cdot <C_{G}(R_{B})>$) So for the protocol to be secure we
want both the above problems to be computationally hard, for the problem A1
to be hard it is required both the centralisers $C_{G}(L_{B})$ and $%
C_{G}(R_{B})$ to be large to resist a brute force type attack.

Attack on $B$'s key in the key exchange protocol. Find an element $%
b_{1}^{\prime }$ $\in L_{B}$ and an element $b_{2}^{\prime }\in R_{B}$ such
that $K_{B}=b_{1}^{\prime }zb_{2}^{\prime }$ where $b_{1}^{\prime
}zb_{2}^{\prime }$ above may be rewritten using a normal form. The pair $%
(b_{1}^{\prime },b_{2}^{\prime })$ is equivalent to the pair $(b_{1},b_{2}),$
because $b_{1}^{\prime }zb_{2}^{\prime }=b_{1}zb_{2}$ this means an attacker
can find the common secret key when this set up is used as part of a key
exchange algorithm as described below.

Then (following \cite{Shpilrain2 et al}) the most obvious way to do the
above attack, Bob private key\medskip

B1. Solve the search version of the membership problem in the double coset $%
<L_{B}>\cdot z\cdot <R_{B}>$ and/or $<L_{B}>\cdot K_{A}\cdot <R_{B}>$%
.\medskip

So for the protocol to be secure we want the above problem to be
computationally hard, for the problem B1 to be hard it is required that the
elements of $L_{B}$,$R_{B}$ cannot all be tested (for an equivalent key,
pair $b_{1}^{\prime },b_{2}^{\prime }$) so to resist a brute force attack.
The above attack can be used to attack the authentication scheme or recover $%
B$'s secret elements with the modification $x=b_{1}^{\prime }z^{\prime
}b_{2}^{\prime }$ instead of $K_{B}=b_{1}^{\prime }zb_{2}^{\prime }$ ($%
x^{\prime }$ is used instead of $K_{B}$ and $z^{\prime }$ is used instead of 
$z$ etc.) so this can be used to impersonate $A$.\medskip\ 

\underline{6.3 Attack When Third Method is used to Choose Subsets in the}

\underline{Variant Authentication Scheme and Variant Key Exchange Scheme}%
\medskip

Attack on $A$'s Key. Find an element $a_{1}^{\prime }$ which commutes with
every element of the subgroup $L_{B}$ and an element $a_{2}^{\prime }\in
R_{B}$ such that $z^{\prime }=a_{1}^{\prime }za_{2}^{\prime }$ where $%
a_{1}^{\prime }za_{2}^{\prime }$ above may be rewritten using a normal form
and $a_{1},a_{2}$ are both invertible elements. The pair $(a_{1}^{\prime
},a_{2}^{\prime })$ is equivalent to the pair $(a_{1},a_{2}),$ because $%
a_{1}^{\prime }za_{2}^{\prime }=a_{1}za_{2}$ this means an attacker can
authenticate as Alice the attack applies to the key exchange protocol with
the modification $K_{A}=a_{1}^{\prime }za_{2}^{\prime }$ instead of $%
z^{\prime }=a_{1}^{\prime }za_{2}^{\prime }$ is solved this gives a
equivalent secret key for $A$ used to get the common secret key by computing 
$\kappa =a_{1}^{\prime -1}K_{B}a_{2}^{\prime -1}$.\smallskip

Then (following \cite{Shpilrain2 et al}) the most obvious way to do the
above attack Alice's private key\medskip

A1. Compute the centraliser of $L_{A},L_{A}\subseteq C_{G}(a_{1})$.

A2. Solve the search version of the membership problem in the double coset $%
<C_{G}(L_{A})>\cdot z\cdot <R_{A}>$ and/or $<C_{G}(L_{A})>\cdot K_{B}\cdot
<R_{A}>$ .\medskip

Attack on $B$'s key in the key exchange protocol is below. Find an element $%
b_{2}^{\prime }$ which commutes with every element of the subgroup $R_{A}$
and an element $b_{1}^{\prime }\in L_{B}$ such that $K_{B}=b_{1}^{\prime
}zb_{2}^{\prime }$ where $b_{1}^{\prime }zb_{2}^{\prime }$ above may be
rewritten using a normal form. The pair $(b_{1}^{\prime },b_{2}^{\prime })$
is equivalent to the pair $(b_{1},b_{2}),$ because $b_{1}^{\prime
}zb_{2}^{\prime }=b_{1}zb_{2}$ this means an attacker can find the common
secret key.

Then (following \cite{Shpilrain2 et al}) the most obvious way to recover Bob
private key\medskip

B1. Compute the centraliser of $R_{A},R_{A}\subseteq C_{G}(b_{2})$.

B2. Solve the search version of the membership problem in the double coset $%
<L_{B}>\cdot z\cdot C_{G}(R_{A})$.$\medskip $

So for the protocol to be secure we want both the above problems to be
computationally hard, for the problem B2 to be hard it is required the
centraliser $C_{G}(R_{A})$ to be large enough to resist a brute force
attack. The key exchange protocol in section 5 has security based upon the
above problem. The above attack can be used to attack the authentication
scheme or recover $B$'s secret elements with the modification $%
x=b_{1}^{\prime }z^{\prime }b_{2}^{\prime }$ instead of $K_{B}=b_{1}^{\prime
}zb_{2}^{\prime }$ is solved ($x$ is used instead of $K_{B}$ and $z^{\prime
} $ is used instead of $z$ etc.) so this can be used to impersonate
Alice.\medskip

\underline{6.4 Requirements of the of Platform Group}\medskip

Hence the platform group $G$ should satisfy at least the following
properties in order for our key establishment protocol to be efficient and
secure, we have taken the requirements from \cite{Shpilrain2 et al} so the
properties are the same as given in \cite{Shpilrain2 et al} with the
relevant modifications. At least one of property P7, P8, or P9 is true
depending on the choice of protocol used.

(P1) $G$ should be a non-commutative group of at least exponential growth.
The latter means that the number of elements of length $n$ in $G$ is at
least exponential in $n$; this is needed to prevent attacks brute force type
attacks on the key space.

(P2) This property may be optional. There should be an efficiently
computable normal form for elements of $G$.

(P3) It should be computationally easy to perform group operations
(multiplication and inversion) on normal forms.

(P4) It should be computationally easy to generate pairs $%
(a,\{a_{1},...,a_{k}\})$ such that $aa_{i}=a_{i}a$ for each $i=1,...,k$.
(Clearly, in this case the subgroup generated by $a_{1},...,a_{k}$
centralizes a).

(P5) For a generic set $\{g_{1},...,g_{k}\}$ of elements of $G$ it should be
difficult to compute $C(g_{1},...,g_{n})=C(g_{1})\cap ...\cap C(g_{k})$.

(P6) This property may be optional. Even if $H=C(g_{1},...,g_{n})$ is
computed, it should be hard to find $x\in H$ and $y\in H_{1}$ (where $H_{1}$
is some fixed subgroup given by a generating set) such that $xwy=w^{^{\prime
}}$, i.e., to solve the membership search problem for a double coset.

(P7) This property may be optional. Even if $H=C(g_{1},...,g_{n})$ is
computed, and $H_{1}=C(g_{1}^{\prime },...,g_{m}^{\prime })$ ($g^{\prime }$
is a generator as usual) is computed and it should be hard to find $x\in H$
and $y\in H_{1}$ such that $xwy=w^{^{\prime }}$, i.e., to solve the
membership search problem for a double coset.

(P8) This property may be optional. Given $H$ and $H_{1}$ is some fixed
subgroups given by a generating sets,and it should be hard to find $x\in H$
and $y\in H_{1}$ such that $xwy=w^{^{\prime }}$, i.e., to solve the
membership search problem for a double coset.

(P9)This property may be optional. Which is there should be an efficiently
algorithm for the word problem in $G$.\medskip

\underline{6.5 Braid groups}\medskip

We now consider braid groups as a possible platform group. Here we consider
the properties (P1)-(P6) from the previous section.

(P1) For $n>2$, braid groups $B_{n}$ are non-commutative groups of
exponential growth.

(P2) There are several known normal forms for elements of B$_{n}$, including
Garside normal form (see \cite{Cha et al}). The Garside normal form is
efficiently computable.

(P3) There are efficient algorithms to multiply or invert normal forms of
elements of $B_{n}$ \cite{Cha et al}.

(P4) It is not so easy to compute the whole centralizer of an element $g$ of 
$G$ \cite{Shpilrain2 et al}. The number of steps required to compute $%
C_{G}(g)$ is proportional the size of the SSS (super summit set) of $g$.
Generally speaking the ``super summit set'' is not of polynomial size in $n$
and the braid length. Nevertheless, there are approaches to finding ``large
parts'' of $C_{G}(g)$, e.g. one can generate a sufficiently large part of $%
SSS(g)$.

(P5) For a generic subgroup $A$ there is no efficient algorithm to compute $%
C_{G}(A)$.

(P6),(P7) and (P8) There is no known solution to the membership search
problem for double cosets $HwH^{\prime }$ in braid groups.

(P9) There are efficient algorithms for the word problem in braid groups,
such as the practical handle reduction algorithm for the word problem
described in \cite{Sibert}.\medskip

\underline{7. Conclusion}\medskip

We have presented new two-pass authentication schemes and key exchange
protocols. This paper is a work in progress, because further work we plan to
do for our authentication scheme is to investigate potential semigroups
(apart from braid groups) and parameters for which it is secure, when its
security is based on the DH-DP or variants of the DH-DP.

\end{document}